# TigerTTS: Draft-Guided Multi-Frame Parallel Speech-Token Language Model for Fast and Robust Text-to-Speech

**Dewei Kong**
Wiz Holdings Pte. Ltd.
Nanjing, China
dewei.kong@wiz.ai, kongdw8@gmail.com

## Abstract

Autoregressive speech-token language models yield high-quality text-to-speech (TTS) synthesis but face three practical bottlenecks: decoding latency scaling with sequence length, word omission or repetition due to the lack of explicit alignment, and audio fidelity limitations imposed by discrete codebooks. We present TigerTTS, a production-grade LLM TTS system built around a lightweight non-autoregressive (NAR) model. Trained with explicit alignment, this NAR model drafts tokens for the entire utterance, which are then refined frame-by-frame by a compact autoregressive LLM. The draft is aligned frame-for-frame with LLM targets and injected at the input, ensuring training-inference consistency. Leveraging the accuracy of this draft, the LLM predicts multiple frames per step, specifically 12 tokens per step across 3 frames and 4 groups. A conditional-total-correlation bound ensures this parallel prediction remains virtually lossless. For tokenization, we employ a codebook-free FSQ tokenizer with four parallel groups and approximately 1800 bps to prevent codebook collapse. The backbone retains positional information exclusively in the input embeddings, enabling exact MLA weight absorption and a four-fold reduction in KV cache size. Decoding is optimized via CUDA Graphs with a graph-safe sampler that preserves the multinomial distribution. Furthermore, we introduce a Fast mode to minimize first-packet latency: the NAR generator acts as a prefix synthesizer with an RTF of 0.03, outputting the first 500 ms of audio before LLM decoding commences. In the default configuration, CUDA-Graph decoding achieves an RTF of 0.05, approximately 20 times real-time, on a single GPU. Despite utilizing a 0.15B backbone, an order of magnitude smaller than competitors, TigerTTS outperforms the standard PyTorch inference of VoxCPM2 (2B) by 5 times and CosyVoice 2 by 16 times. Empirically, TigerTTS achieves the lowest CER and WER on the SEED sets (1.09 / 1.67) and superior WER, NMOS, and speaker similarity on LibriSpeech test-clean (2.22 / 4.36 / 0.820). Ultimately, the guided semantic tokens from the Fast-mode generator provide a well-aligned content prior, effectively eliminating dropped words, loops, and runaway generation without compromising inference speed.

## 1 Introduction

Recent text-to-speech work splits, roughly, into two lines. The first is non-autoregressive (NAR) acoustic models—FastSpeech, Glow-TTS, Matcha-TTS, StyleTTS—where phonemes are stretched into frames by an explicit duration or a monotonic alignment and the mel-spectrogram is generated in parallel. These systems are quick and they do not skip or repeat words. The cost is that a $L_1/L_2$ regression objective only recovers the conditional mean (or median) of the data, so the output comes out over-smoothed and the prosody flat. The other line is discrete-token language-model TTS, VALL-E and the CosyVoice family, which models $p(token|text)$ autoregressively and handles the one-to-many text-to-speech mapping far better. Here the trade-off runs the other way: decoding token by token is slow, and with no explicit alignment it is precisely the long sentences, rare words and digit strings that lose or repeat a word, stop too early, or run past the end.

TigerTTS keeps the language model but gives it a different job. A small NAR draft model, first proposes draft tokens for the whole utterance in the same discrete space as the final targets, with alignment and length already settled. The LLM then refines the draft autoregressively, and at step $k$ it already has in view the drafts of the very $R$ frames it is about to predict. Note that there is no accept/reject stage as in speculative decoding: the LLM always re-samples with the draft as a condition, so the draft only guides generation and we do not claim any lossless speed-up. The list that follows summarises what we contribute; together the pieces cover the full pipeline—text-side emotion/prosody prediction → ASR forced alignment → NAR draft generator → FSQ tokenizer → draft-guided LLM → mel decoder → speaker-conditioned Vocos—along with the losses and derivations for each stage, and Figure 1 sketches the inference path. The rest of the paper is laid out as: §2 related work, §3 method, §4 training recipe, §5 evaluation, §6 limitations, §7 conclusion.

- Frame-aligned draft guidance. Draft tokens line up frame-for-frame with the LLM's targets: when the LLM predicts frames $kR, \dots, kR+R-1$, it is looking at the drafts for exactly those frames. We give the alignment rules and show the draft inputs match between training and inference (Prop. 2). The draft features enter only once, at the LLM input, concatenated with the audio embedding and the speaker vector before the backbone. These guided tokens are the heart of Fast mode: the prior they carry caps the LLM's uncertainty, which is why decoding stays steady in places where other LLM TTS systems drop words or loop.

- Multi-frame parallel prediction, with a reason for it. Each LLM step emits $R = 3$ frames by $\times\ G = 4$ groups, 12 tokens in all, cutting the sequence to $1/R$. Via the conditional total correlation we show that the modelling error from assuming the tokens within a step are independent is bounded by the sum of their conditional entropies, and the draft pushes those toward zero, so a better draft makes the parallel step closer to lossless ( §3.5).
- A codebook-free parallel speech tokenizer. FSQ takes HuBERT features as input and outputs 4 parallel groups × 512 classes per 20 ms frame (about 1800 bps), so there is no codebook that could collapse. We derive the shift boundary for even levels, show the origin is a fixed point with exactly $L$ grid points, and give the mixed-radix encoding with its exact inverse ( §3.2).
- A backbone chosen with deployment in mind. RoPE touches only the input embeddings and never the attention, so MLA weight absorption holds exactly; no separate $q_{pe}/k_{pe}$ branch is required and the KV cache is compressed 4 × (640 against 2560 floats per position per layer; §3.6).
- A single CUDA-Graph decode with an exact sampler. The draft- and speaker-dependent linear terms are factorised block-wise and computed in advance, so the whole single-step decode (embedding, absorbed MLA layers, output head, sampling, end detection, buffer write-back) sits in one CUDA Graph and runs on the device; the graph-safe exponential sampler it uses has the same distribution as multinomial sampling ( §3.7).
- Fast mode, for first-packet streaming. The draft generator uses the LLM's own token vocabulary and runs a fully parallel encoder–decoder–Vocos path at an RTF of 0.03, so it can output the first 500 ms of audio before the LLM has decoded a token; the LLM then picks up from those tokens as real history ( §3.8).

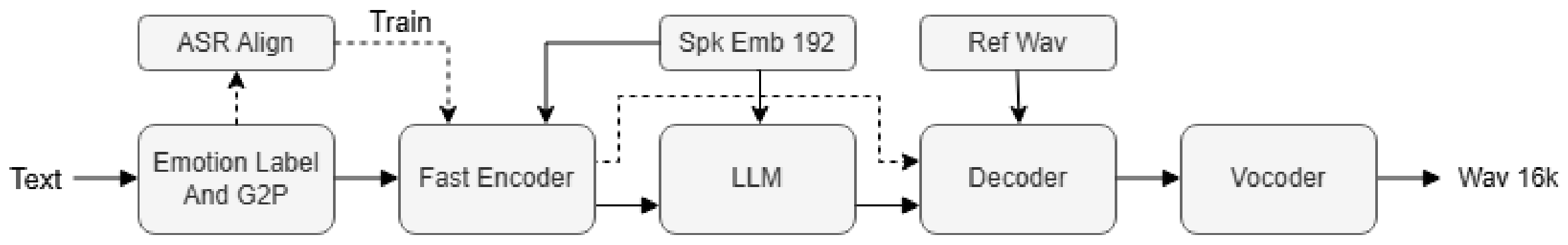

Figure 1. An overview of the TigerTTS inference pipeline.

## 2 Related Work

Language-model TTS. VALL-E models the RVQ tokens of EnCodec with an AR Transformer; CosyVoice 2 swaps VQ for FSQ as its semantic tokenizer and gets better codebook use. TigerTTS also uses FSQ, but as $G = 4$ parallel groups rather than a single codebook, and its LLM predicts several frames and groups at each step under frame-aligned draft conditioning. RQ-Transformer and Moshi predict multiple levels at a position through a depth transformer; here we use one linear head over all $R \times G$ tokens and let the draft keep the intra-step independence assumption harmless. VoxCPM, by contrast, moves to a hierarchical continuous-latent diffusion-AR design and drops discrete tokens. TigerTTS stays with discrete FSQ tokens, because it is their being draftable that makes frame-aligned draft guidance possible.

**Speculative decoding and draft models. TigerTTS uses a draft-plus-main-model structure, but treats the draft as a standing condition rather than a proposal to be checked; no candidate is accepted or rejected.**

**Streaming.** In the Fast mode of §3.8 the draft also serves as a prefix synthesiser, so the first packet comes from the draft generator itself, in the LLM’s own token space; CosyVoice 2 takes a different route to streaming, through a chunk-aware flow-matching redesign in a separate path.

**Efficient attention.** DeepSeek-V2’s MLA compresses the KV cache into a low-rank latent and reads it back through weight absorption; because TigerTTS keeps positional information out of attention, absorption holds exactly, which removes the need for a separate positional branch.

**Vocoders.** Vocos [13] predicts STFT magnitude and phase with a ConvNeXt backbone and applies ISTFT; TigerTTS adds a 192-d ECAPA speaker embedding at its input.

## 3 Method

### 3.1 Overview and notation

Figure 1 provides the system-level overview of the whole inference pipeline, while Figure 2 zooms in on the LLM architecture at its core. The text is first passed through a text-side emotion/prosody tagger that predicts discrete emotion and prosody tags (seven emotion categories, five speaking styles and phrase breaks graded #1–#4); these tags are placed in the LLM's flag prefix (§3.4). Text is normalised and converted to phoneme ids $x \in N^{T_p}$ by G2P; the speaker is represented twice: an ECAPA embedding $e \in R^{192}$, used by the draft generator, the LLM and the vocoder, and a global style vector $g_e \in R^{1024}$ extracted offline from a reference mel by a MelStyleEncoder, used by the decoder. The whole system shares one set of log-mel features: 16 kHz, $n_{fft} = 1024$, window 640 (40 ms, Hann), hop 160 (10 ms), 100 mel bins, dynamic-range compression $Y = lnmax(M|X|, 10^{-5})$ and no mean/variance normalization; frame counts are truncated to an even $T_m$, so mel frames and 20 ms token frames stand in a strict $2:1$ relation.

The audio token sequence is $a = [\langle s\rangle, a_1, \dots, a_n, \langle e\rangle, \langle stop\rangle]$, with $a_t \in [0, K)^G$. The prediction target is $y = a_{1:M}$; it is cut into chunks of $R$ frames, $y = [c_0, \dots, c_{S-1}]$ with $c_k = (y_{kR}, \dots, y_{kR+R-1})$ and $S = \lceil M/R \rceil$, the remainder being right-padded by $\langle stop\rangle$. TigerTTS models

$$p_\theta(y|x, e, g) = \prod_{k=0}^{S-1} p_\theta(c_k|c_{<k}, x, e, g), \qquad g_k = (g_{kR}, \dots, g_{kR+R-1}), \tag{1}$$

where $g$ is the NAR draft aligned by the rules of §3.4. Within a chunk a further conditional-independence factorisation is used,

$$p_\theta(c_k|\cdot) = \prod_{r=0}^{R-1} \prod_{q=1}^{G} p_\theta((q)_y^{kR+r}|h_k), \tag{2}$$

and $h_k$ is the hidden state produced by the LLM at step $k$; one forward pass therefore emits $R \times G = 12$ tokens. §3.5 analyses when this factorisation is lossless.

Three time resolutions coexist: 50 Hz audio tokens (20 ms), 100 Hz mel frames (10 ms) and 16 kHz samples. One second of audio corresponds to 50 token frames (200 tokens), ≈16.7 LLM steps, 100 mel frames and 16000 samples; 16 s of speech occupies only ≈267 of the $L_{max} = 1024$ LLM positions.

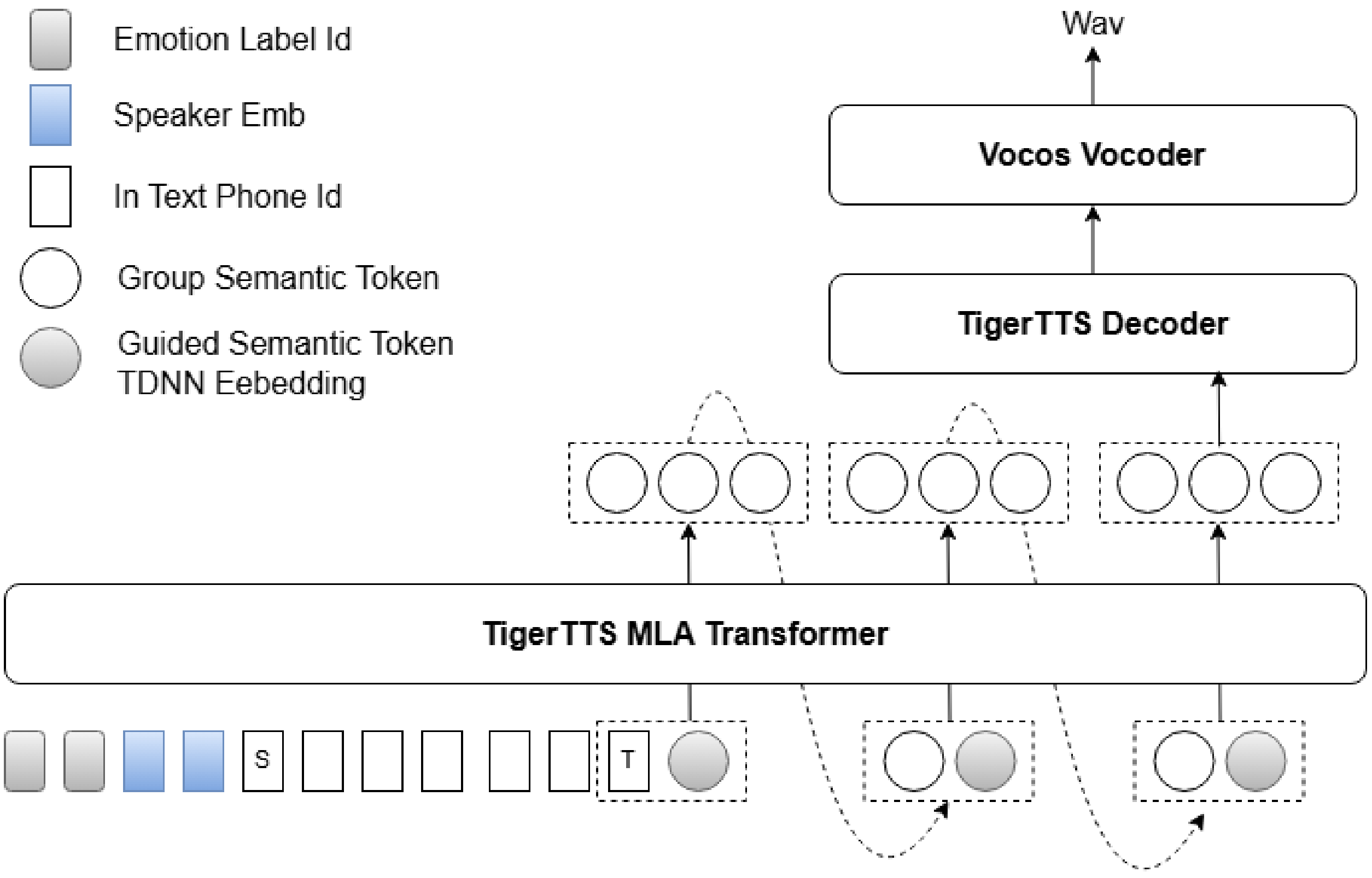

Figure 2: The TigerTTS pipeline.

### 3.2 Codebook-free parallel speech tokenizer (FSQ)

The tokenizer maps HuBERT features to discrete tokens without any codebook. Each of the $G = 4$ groups takes $d = 4$ real dimensions and rounds them onto $L_j$ levels with $L = [8,4,4,4]$, so each group has $K = \prod_j L_j = 512$ classes and a frame carries $\sum_j log_2 L_j = 36$ bits, for a nominal bitrate of $36 \times 50 = 1800$ bps. Because every dimension is bounded independently, no vector-quantiser machinery is needed, whether commitment loss, EMA updates or dead-code restarts, and there is no codebook to collapse.

Bounding with a shift. To make rounding exact, $z_j$ is squashed by the odd-symmetric bounded transform

$$b_j(z) = \eta_j tanh(z + s_j) - o_j, \qquad \eta_j = \frac{(L_j-1)(1-\varepsilon)}{2}, \quad o_j = \begin{cases} \frac{1}{2}, & L_j \ even, \\ 0, & L_j \ odd, \end{cases} \quad s_j = artanh\frac{o_j}{\eta_j}, \tag{3}$$

with $\varepsilon = 10^{-3}$. Two properties follow. (i) The origin is a fixed point: $b_j(0) = \eta_j tanh(s_j) - o_j = 0$, so at initialisation, where $z$ is centred at zero, tokens land exactly on grid centres instead of being pushed to one side. (ii) Rounding yields exactly $L_j$ integers: for even $L_j$ the image tends to $(-L_j/2, L_j/2 - 1)$ and rounds onto $\{-L_j/2, \ldots, L_j/2 - 1\}$; for odd $L_j$ it is $(-(L_j-1)/2, (L_j-1)/2)$ and rounds onto $\{-(L_j-1)/2, \ldots, (L_j-1)/2\}$. The role of $\varepsilon$ is to keep the endpoints strictly inside the outermost rounding ranges. Levels below 3 are impossible: for $L_j = 2$ one has $o_j/\eta_j > 1$ and $artanh$ is undefined, which is why binary quantisation requires methods such as LFQ. After rounding, values are normalised by the half-width $w_j = \lfloor L_j/2 \rfloor$ into $[-1,1]$.

Gradients flow through the straight-through estimator (STE); writing $q_j = round(b_j(z_j))/w_j \in [-1,1]$ for the normalised quantised value, we have

$$round_{ste}(y) = y + sg[round(y) - y], \qquad \frac{\partial q_j}{\partial z_j} = \frac{1}{w_j}\eta_j\left(1 - tanh^2(z_j + s_j)\right), \tag{4}$$

where $sg$ is stop-gradient; all supervision comes from the downstream mel reconstruction loss. When $|z_j|$ is large the $tanh$ saturates and the gradient vanishes, so the input projection is scaled such that $Std(z) \approx 1.25$, where source-code measurements show the highest token perplexity. Utilisation is monitored by the per-group perplexity $PPL = exp(-\sum_c plogp) \le K$, i.e. at most $log_2 K = 9$ bits.

Mixed-radix encoding makes the groups invertible. With

$$\beta_1 = 1, \qquad \beta_j = \prod_{i<j} L_i, \qquad c^{(q)} = \sum_{j=1}^{d} k_j \beta_j \in \{0, \ldots, K-1\}, \tag{5}$$

the code $c^{(q)}$ (for $L = [8,4,4,4]$, $\beta = [1,8,32,128]$) is decoded exactly by $k_j = \lfloor c/\beta_j \rfloor mod L_j$, so $decode(encode(z))$ is bit-wise identical to the forward quantiser output and the decoder at inference follows the same numerical path as during training. Compared with residual vector quantisation, parallel groups have no residual chain: all groups are predicted symmetrically and simultaneously by the LLM, one step per frame chunk.

### 3.3 Draft generator: the non-autoregressive encoder

Stage A is a FastSpeech-style NAR encoder that predicts in the same discrete space as the tokenizer. A phoneme embedding (508-d, unit-variance initialisation) concatenated with a 4-d compressed speaker vector is encoded by a 4-layer bidirectional Transformer (512 wide) with RMSNorm, QK-Norm + RoPE, a Conformer convolution branch (kernel 31) and SwiGLU (1408 hidden). QK-Norm bounds the attention logits by learnable gains and prevents softmax saturation; the convolution branch re-applies its mask after every bias-carrying operation so that padding cannot leak into valid frames. A dilated duration predictor (dilation 1, 2, 4, 1) outputs log-durations in 20 ms units, and length regulation copies each phoneme vector $d_i$ times by cumulative-sum differencing.

Duration rounding at inference uses cumulative rounding,

$$C_k = round\left(\sum_{i\le k} w_i\right), \qquad d_k = C_k - C_{k-1}, \qquad w_i = \alpha\, e^{logw_i}, \tag{6}$$

where $\alpha$ is the speed factor. Proposition 1 (cumulative rounding bounds phoneme boundaries). $|\sum_{i\le k} d_i - \sum_{i\le k} w_i| \le \frac{1}{2}$ for every prefix, by telescoping. Ceil-rounding instead adds 0.5 frames per phoneme on average ($\approx$ 1 s extra for $T_p = 100$), and per-phoneme rounding accumulates errors up to $T_p/2$ at the sentence end; cumulative rounding keeps every phoneme boundary within half a frame (10 ms), and $max(d_k, 1)$ guarantees one frame per phoneme.

A residual decoder with AdaIN blocks (speaker-conditioned) produces a 100-d mel draft at 10 ms after a $\times$ 2 upsample, and an affine token head converts the mel draft into logits over $G \times K$ classes; the head has no nonlinearity between two adjacent mel frames, which forces the encoder to organise the token-discriminative information into the mel draft itself. Training combines a log-domain MSE duration loss (the maximum-likelihood fit for log-normal durations), an $L_1$ mel loss (a Laplace likelihood, robust to outliers), and a cross-entropy token loss against the FSQ quantisation of real HuBERT features (random guessing gives $ln512 \approx 6.24$); the token weight is linearly ramped from step 4000 to 100000 while the mel weight decays to a 0.01 floor.

### 3.4 Draft-guided LLM: alignment, features, and one injection

Sequence layout. The LLM input is a single sequence of $10 + T_p + S \le 1024$ positions: a 10-frame prefix (8 flag frames carrying the emotion and prosody tags predicted from the text by the text-side tagger of §3.1, plus 2 voiceprint frames), the text, and $S$ audio steps. The prefix is projected to 320 dims and tiled $4 \times$ to $H = 1280$; it receives no positional encoding. Each step-$k$ input $\xi_k$ concatenates three parts: the embeddings of the previous chunk $c_{k-1}$ ($R$ copies of $\langle s \rangle$ when $k = 0$) with input-side RoPE by frame index; the $R$ draft feature frames $g_k$; and a 4-d speaker projection repeated at every step, a direct speaker condition far from the prefix that mitigates timbre drift on long sentences.

Frame alignment. The draft $g \in [0,512)^{4\times m}$ is aligned to any output length $N$ by

$$g_t = \begin{cases} g_{t+1}, & 0 \le t < m, \\ \langle e_g \rangle, & t = m, \\ \langle pad_g \rangle, & m < t < N, \end{cases} \tag{7}$$

and is simply truncated when $m \ge N$; $\langle e_g \rangle$ and $\langle pad_g \rangle$ are dedicated draft tokens whose ids are disjoint from the audio's. The draft and the target are co-indexed frame-for-frame: when predicting a particular frame the LLM sees the draft of that very frame. This differs from ordinary prefix conditioning, where the model has to infer its position in the conditioning sequence on its own; here the NAR encoder resolves that alignment. The draft end marker $\langle e_g \rangle$ sits exactly at the frame where the draft believes the audio ends, next to the true $\langle e \rangle$, providing a strong end prior.

Length alignment. At training, with $n_a$ the padded target length and $m$ the draft length, $l^\star = max(n_a, m)$: the audio is right-padded with $\langle stop \rangle$ to $l^\star$ frames. The padded stops take part in the loss, so the model learns to emit stop even where the draft thinks the utterance continues; the draft is then aligned to $l^\star + 1$ effective frames, so that the $\langle e_g \rangle$ at index $l^\star$ survives instead of being erased into padding. At inference the cap is

$$N_{max} = min\Big(\lfloor 1.5m \rfloor + 50, (L_{max} - 10 - T_p)\, R\Big), \tag{8}$$

and the draft is aligned to $N_{max}$.

Proposition 2 (training/inference consistency). For every frame $t < min(l^\star, N_{max})$, the draft tokens seen by the LLM are identical in training and inference. Proof: both sides call the same align function, whose value at $t$ depends only on the relation between $t$ and $m$, and padding touches only frames at or beyond the effective length. □ The training drafts are produced by running the same, already-trained encoder on the training texts and caching its output. Ground-truth tokens are never corrupted, so training and inference drafts share one distribution and there is no exposure bias.

Draft features. Each draft frame is embedded with a lookup table and passed, frame-level, through a non-causal TDNN with expansion ratio $\rho = 2$ in its 1×1 projection. A depthwise convolution with kernel 21 gives a $\pm 200$ ms receptive field with one temporal filter per channel, followed by a $1 \times 1$ expand–GELU–contract block, an identity residual, and a layer scale initialised at $10^{-2}$. The mask is applied after LayerNorm and before the convolution; otherwise padding frames become the (nonzero) LN bias and leak through the depthwise filter into valid frames near the boundary. The per-frame draft feature is

$$\Gamma_t = \big[\varepsilon_t;\ TDNN(\varepsilon)_t\big] \in R^{2048}, \qquad \Gamma_k = [\Gamma_{kR}; \dots; \Gamma_{kR+R-1}] \in R^{RD_f}, \tag{9}$$

with non-overlapping per-step windows. Because the draft is a NAR output fully known in advance, the non-causal TDNN is computed once over the whole draft at inference and sliced per step, exactly as in training, and costs no streaming latency.

One injection site. The input side projects the concatenation of audio embedding, draft feature and speaker vector,

$$\xi_k = W_1[\alpha_k;\ \Gamma_k;\ \kappa], \qquad W_1 \in R^{1280\times 9988}, \qquad \kappa = W_\kappa e + b_k \in R^4, \tag{10}$$

so that the draft takes part in the 6-layer Transformer together with the history tokens and the text. A single linear head then emits all R×G logit blocks of a step at once.

### 3.5 Why the draft makes multi-frame parallel prediction work

Proposition 3 (multi-frame error equals the conditional total correlation). Let $C = (c_{<k}, x, e, g_k)$ collect all conditions. The error introduced by replacing the joint $p(c_k|C)$ with the product $\prod_i p(y_i|C)$ is exactly the conditional total correlation,

$$KL(p(c_k|C) \parallel \textstyle\prod_i p(y_i|C)) = \sum_i H(y_i|C) - H(c_k|C) \ \triangleq\ TC(c_k|C), \tag{11}$$

and since $H(c_k|C) \ge 0$,

$$0 \le TC(c_k|C) \le \textstyle\sum_{i=1}^{RG} H(y_i|C). \tag{12}$$

Corollary: if the draft $g_k$ carries more information about $c_k$, each conditional entropy $H(y_i|C)$ decreases and the bound tightens; if the draft determines $c_k$, then $TC = 0$ and the conditional-independence assumption is lossless. Information never decreases when conditions are added, so the draft-guided bound is never worse than multi-frame prediction without a draft. The same mechanism allows TigerTTS to emit $R = 3$ frames per step: the strong correlations among frames within a phoneme and among groups within a frame are largely accounted for by the draft, and what is left for the LLM to resolve carries little entropy and is close to independent.

### 3.6 Efficient backbone: input-only RoPE and exact MLA absorption

The backbone is a 6-layer pre-norm Transformer with $H = 1280$, 10 heads × 128 dims and a dense SwiGLU with $d_{ff} = 3456$; residual output projections are scaled by $1/\sqrt{2L}$. Positional information lives only on the input embeddings: RoPE with $\Theta = 3000$ is applied to text from position 0 and to audio by frame index: the $r$-th frame of step $k$ gets position $kR + r$, so the $R$ frames of a step carry distinct phases while consecutive steps differ by multiples of $R\theta_i$. The prefix and the draft receive none, the draft being located implicitly by its frame-wise concatenation with audio. Applied this way RoPE is an absolute, not relative, encoding: in general $top_W^q W_k$ does not commute with the rotation, so the attention logit cannot be reduced to a function of $m - n$. Counting by frame index rather than step index is what keeps the $R$ frames within a step distinguishable.

This choice has a precise payoff. The MLA projects $u_t$ to a query and to a low-rank latent that constitutes the entire KV cache,

$$\lambda_t = RMSNorm(W_{DKV} u_t) \in R^{r_{kv}}, \qquad (i)_k^t = (i)_W^{UK} \lambda_t, \qquad (i)_v^t = (i)_W^{UV} \lambda_t, \qquad i = 1, \dots, h, \tag{13}$$

and during decoding the keys fold into the query while the values fold out of the output,

$$(i)top_q^t (i)_k^j = \left((i)top_W^{UK} (i)_q^t\right)^{top} \lambda_j, \qquad (i)_o^t = (i)_W^{UV} \left(\textstyle\sum_j a_{tj} \lambda_j\right). \tag{14}$$

All heads therefore share a single $r_{kv}$-dimensional latent head, and one attention call serves all of them. The derivation requires $k_j$ to be a linear function of $\lambda_j$ alone; any RoPE inside attention would introduce $top_R^t R_j$ and break precomputation. With positional encoding kept out of attention, absorption is exact and needs no separate $q_{pe}/k_{pe}$ branch. The cache per layer and position is $r_{kv} = 640$ vs. $2hd_n = 2 \times 10 \times 128 = 2560$ floats, a $4\times$ compression ($\approx$ 15.7 MB in fp32 at full length), which at small batch translates directly into decode speed because decoding is memory-bandwidth-bound.

### 3.7 Decode-time design: legality, sampling, CUDA Graph

Legality and stopping. A logit bias restricts every group to legal codec tokens ($v < K$) and permits end tokens only in group 1, so the stop decision is concentrated in one place and the decoder input is guaranteed legal. End detection is per frame within a step (cumulative sigmoid over the $R$ frames), so generation can stop at any frame, not only at step boundaries; frames after the first end are zeroed and excluded from the output length. If the very first frame ends (output length 0), the system falls back to the draft tokens themselves, so the decoder never receives an empty sequence. Reaching $N_{max}$ without ending triggers truncation and a warning.

Sampling. Temperature, top-$k$ and top-$p$ are applied per token; top-$p$ uses an exclusive cumulative sum so that the most probable token is always kept. The deployment configuration is $T = 1.0$, top-$k = 5$, top-$p = 0.9$, with the seed reset to 1234 per request. $torch.multinomial$ synchronises GPU and CPU and cannot be captured in a CUDA Graph; TigerTTS instead samples with a single reduction,

$$v = argmax_v \frac{p_v}{\varepsilon_v}, \qquad \varepsilon_v \sim Exp(1). \tag{15}$$

Proposition 4 (graph-safe exponential sampling is exact). $Pr(v = \hat{v}) = p_{\hat{v}}$. Proof: $X_v = \varepsilon_v / p_v \sim Exp(p_v)$ independently, so $argmax_v p_v/\varepsilon_v = argmin_v X_v$ and for independent exponentials

$$Pr(X_{\hat{v}} = min_v X_v) = \int_0^\infty p_{\hat{v}} e^{-p_{\hat{v}} x} \prod_{v \neq \hat{v}} e^{-p_v x}\, dx = \frac{p_{\hat{v}}}{\sum_v p_v} = p_{\hat{v}}. \quad \square \tag{16}$$

Tokens cut by top-$k$/top-$p$ have $p_v = 0$ and are never selected; the trick is equivalent to the Gumbel-max transform and can be captured in the graph, which multinomial sampling cannot.

CUDA Graph. The step input factorises block-wise,

$$\xi_k = a_W^1 \alpha_k + g_W^1 \Gamma_k + s_W^1 \kappa, \tag{17}$$

where the draft and speaker terms do not depend on the generated output and are precomputed for all steps into a static buffer; each step then reduces to an $H \times RH$ product plus the absorbed attention. The single-step decode (embedding, 6 MLA layers, output head, sampling, end detection, buffer write-back) is captured in one CUDA Graph: a static KV cache of length $L_{max}$ with unwritten positions masked by $1[j \leq kv_pos]$; step counters, done flags and outputs held in GPU tensors updated in place; completion checked only every 16 replays; graphs cached per (batch, $T$, top-$k$, top-$p$) and warmed up at load. Under greedy decoding the graph version is bit-identical to eager decoding, which again matches the argmax of the teacher-forcing forward on the same inputs (a source self-check requires accuracy above 0.8).

Complexity. The only serial stage of the whole system is LLM decoding, with

$$S = \left\lceil \frac{T_a}{R} \right\rceil \approx 16.7 \; steps\ per\ second\ of\ audio, \tag{18}$$

that is, $1/3$ of frame-by-frame ($R = 1$) autoregression and $1/12$ of flat token-by-token modelling; the quadratic attention term shrinks by $1/R^2$. Table 1 summarises each stage.

Table 1: Per-stage computational cost. Only LLM decoding is serial.

| Stage | Structure | Cost per stage | Parallel? |
|---|---|---|---|
| Draft generator | 4-layer bidirectional Transformer + convs; 28.8M params | $O(2_T^p d + T_g d^2)$ | fully parallel |
| LLM prefill | 6-layer causal attention; 145.7M params | $O(L(10 + T_p)^2 H)$ | parallel |
| LLM decode | 1 position per step | $O(L(H^2 + pos \cdot r_{kv} h)) \times S$ | serial, $S$ steps |
| Decoder | Transformer + convs; 139.8M params | $O(2_T^a d + T_m d^2)$ | fully parallel |
| Vocoder | 12-layer ConvNeXt + ISTFT; 40.3M params | $O(T_m d^2 + T_m n_{fft} log n_{fft})$ | fully parallel |

### 3.8 Fast mode: first-packet streaming by joint synthesis with the draft generator as prefix synthesiser

Interactive systems tend to be judged on first-packet latency—how long it takes from receiving the text to the first playable audio—and §5.2 reports it next to RTF. In this design that number falls straight out of how the models are arranged. The draft generator is not only a conditioning signal: it works in the same token space as the LLM, since its affine token head gives logits over the very G×K classes the LLM targets, and it already carries a full, fully parallel route to a waveform (phoneme encoder, duration predictor, mel-draft decoder, Vocos) that runs at an RTF of 0.03 on our deployment GPU. That lets it stand on its own as a prefix synthesiser, and we call this joint-synthesis setup the Fast mode.

Fast mode moves along a single continuous token stream. It opens with the first packet, produced on the Fast-mode path: Stage A runs by itself and immediately emits the first P frames, whose mel draft goes through Vocos and streams out before the LLM has taken a step. We take P as a multiple of R—for a three-second utterance, P = 24 frames (480 ms) or P = 75 frames (1.5 s). At RTF 0.03 the prefix costs about P/50 × 0.03 seconds of GPU time, roughly 15 ms for 500 ms of audio, so the first packet keeps pace with the text front-end. From there the LLM carries the continuation: the draft tokens for the prefix are fed in as real audio tokens, sitting exactly in the history slots the step inputs reserve for earlier chunks, and the LLM resumes decoding at the boundary step, producing the rest of the utterance and streaming it chunk by chunk as in the default path. The prefix tokens are simply history once they have been emitted.

Timbre stays put across the boundary, with no cross-fading needed. Both phases share the same speaker conditioning (the ECAPA embedding feeds the draft generator, the LLM and the vocoder alike) and the same Vocos instance, and the continuation is conditioned on the draft generator's real tokens, so the LLM keeps up the prosody the listener has already heard; the frame-aligned draft guidance—the whole draft still injected at the one input site—keeps the continuation tied to the same alignment. The one distributional mismatch is that the history prefix holds draft tokens where the default path would hold LLM outputs. A short fine-tune removes it: the ground-truth history of the first P/R steps is swapped for cached draft tokens (the same cached-draft pipeline as §4), after which prefix-conditioned continuation is in-distribution.

Not many speech-token LLM systems can do this. An ordinary system has just one model that produces acoustic tokens—the LLM itself—so its first packet must either wait for the LLM's serial decode (roughly 2.88–60× slower per second of audio, Table 4) or fall back on a separately trained streaming path in a different token space, the chunk-aware redesign CosyVoice 2 built streaming around, at an RTF of 0.8. In TigerTTS the two roles are split from the beginning: a 28.8M fully parallel NAR model that shares the LLM's vocabulary, and a 145.7M LLM trained to condition on that model's tokens. Fast mode is a direct consequence—the same property that gives us draft guidance gives us the prefix. It needs no more than the same models plus one extra fine-tune; the architecture does not change.

## 4 Training Recipe

Training runs in stages that depend on each other. (0) A text-side tagger is trained on annotated text to predict discrete emotion and prosody labels, which fill the flag prefix at inference (§3.1). (1) An ASR aligner—CTC plus attention with position-sensitive monotonic attention, sharpened by monotonic alignment search—gives phoneme duration labels. (2) The decoder learns to reconstruct mel from FSQ tokens of real HuBERT features, and that same FSQ quantiser is then used to tokenise the whole data set. (3) The draft generator is trained with duration, mel and token objectives, and its output on the training texts is cached as drafts. (4) The LLM is trained with teacher forcing on the cached drafts. One short extra stage fine-tunes the LLM for Fast mode (§3.8), replacing the ground-truth history of the first P/R steps with the cached draft tokens so that prefix-conditioned continuation is in-distribution. The deployed decoder is the same checkpoint used to extract tokens, which keeps token semantics consistent. The LLM uses AdamW ($\beta = (0.9, 0.95)$, weight decay 0.1 on matrices only, gradient clip 1.0) at a learning rate of $2 \times 10^{-4}$, a 2000-step warmup and cosine decay to 10%, batch 16. Table 2 lists the main hyper-parameters.

Table 2: Main hyper-parameters (condensed).

| Component | Setting |
|---|---|
| FSQ tokenizer | $G = 4$ groups $\times$ $K = 512$ (levels $[8,4,4,4]$); 20 ms frames; nominal 1800 bps |
| Draft generator | 4 layers × 512; SwiGLU 1408; duration predictor dilation (1, 2, 4, 1); decoder width 384 |
| LLM backbone | $H = 1280$, 6 layers, 10 heads × 128; $r_q = r_{kv} = 640$; $d_{ff} = 3456$; $R = 3$; $L_{max} = 1024$; $\Theta = 3000$ |
| Special tokens | audio $\langle s \rangle = 512$, $\langle e \rangle / \langle stop \rangle = 513$; draft $\langle e_g \rangle / \langle pad_g \rangle = 514$ (vocab $520$) |
| Prefix | 8 flag slots (predicted emotion/prosody tags) + 2 voiceprint frames; no positional encoding |
| Optimisation | AdamW (0.9, 0.95); wd 0.1; clip 1.0; lr $2 \times 10^{-4}$, warmup 2000, cosine → 10%; batch 16 |
| Inference | $T = 1.0$, top-$k$ 5, top-$p$ 0.9; seed 1234; $N_{max} = \lfloor 1.5m \rfloor + 50$ |

# 5 Experiments

## 5.1 Tokenizer Evaluations

A good speech tokenizer should use its codebook fully, retain information faithfully, and be independent of the speaker. TigerTTS quantizes HuBERT features with a group-parallel FSQ: the resulting token tensor has shape [batch, 4, frame_num], where each of the 4 groups takes integer values in [0, 512). Following the tokenizer evaluation protocol of the CosyVoice 2 study, Table 3 reports the codebook utilization and the ASR error rate within the entire encoder. The group-parallel FSQ uses all 2,048 codes (4 groups ×512) and yields an ASR error rate about 36–43% lower than the FSQ tokenizer of the CosyVoice 2 study on all four benchmarks (CommonVoice and Fleurs, English and Chinese), which suggests that quantizing HuBERT features retains more semantic information.

Table 3: The comparison of VQ and FSQ inside the SenseVoice-large encoder, following the evaluation protocol of the CosyVoice 2 tokenizer study; the TigerTTS row is the group-parallel FSQ tokenizer over HuBERT features (4 groups ×512 codes). C.V. stands for the CommonVoice benchmarks. The best ASR error rate in each column is in bold.

| Method | Codebook Size | Util. | ASR Error Rate (%) | | | |
|---|---|---|---|---|---|---|
| | | | C.V. EN | C.V. CN | Fleurs EN | Fleurs CN |
| VQ | 4,096 | 963 (23%) | 18.26 | 11.56 | 7.65 | 5.03 |
| FSQ | 6,561 | 6,561 (100%) | 10.67 | 7.29 | 6.58 | 4.43 |
| **TigerTTS (ours)** | **2,048** | **2,048 (100%)** | **6.35** | **4.66** | **4.19** | **2.52** |

## 5.2 Evaluation protocol

Intelligibility is measured by transcribing synthesised speech with an ASR model and reporting WER/CER, counting skipped words, repetitions and non-termination (hitting $N_{max}$) separately; speaker similarity by the cosine similarity of ECAPA embeddings; naturalness by N-MOS; efficiency by RTF (wall-clock generation time divided by output-audio duration), first-packet latency (minimised by the Fast mode of §3.8) and GPU memory. Ablations: (a) remove the draft (all $\langle pad_g \rangle$); (b) remove the draft feature at the input; (c) $R \in \{1,3,5,7\}$; (d) remove the draft TDNN; (e) feed draft tokens directly to the decoder, skipping the LLM, as a lower bound. Draft-quality analysis measures the frame-and-group agreement between LLM outputs and the draft, and plots WER across draft-accuracy buckets, to validate §3.5.

Model size. The deployed system comprises four modules totalling 354.6M parameters, i.e. 0.35B; the counts are obtained by loading each checkpoint and summing its tensor elements, and all weights are float32, so file size divided by four bytes reproduces the same number. Stage A (draft generator) holds 28.8M parameters: Transformer encoder 21.7M, duration predictor 3.4M, mel decoder 3.1M and FSQ token head 0.7M. Stage B (LLM) holds 145.7M: Transformer layers 115.5M, the draft-feature splice 12.8M, the draft TDNN 4.2M, the audio-token output head 8.0M, and about 5.2M of embeddings. Stage C (decoder) holds 139.8M: text encoder 38.1M, the three residual convolution blocks 72.4M, BERT-style layers 16.6M, reference speaker encoder 8.5M, the remainder being projections. Stage D (Vocos vocoder) holds 40.3M: ConvNeXt backbone 39.3M and ISTFT head 1.1M. The FSQ codebook itself carries no learned parameters.

## 5.3 Inference speed

Table 4 compares decoding speed. Every RTF in it was measured by us on the same RTX 4090 (24 GB), single stream; none is lifted from a source paper or converted from different hardware. TigerTTS is run in its default setting of three frames per step with the CUDA-Graph single-step decode at batch = 1. It decodes 5 × faster than the PyTorch serving path of VoxCPM2 and 2.6 × faster than that model's optimised Nano-vLLM path, about 16 × faster than CosyVoice 2, and 2.88 × faster than the 0.8B VoxCPM1.5, even though its 145.7M LLM backbone is an order of magnitude smaller than the 2B VoxCPM2. Several factors explain the gap: the sequence is shortened 3 × by multi-frame parallel prediction, the 4 ×-compressed KV cache is read in a single absorbed attention call over a 640-d latent, and each step is one CUDA-Graph replay with no host round-trip.

Table 4: Inference speed on a single RTX 4090 (RTF, lower is better; batch 1, TigerTTS uses the CUDA-Graph decode). All RTFs were measured by us on the same RTX 4090 (24 GB) in a single-stream setting; none is quoted from a source paper or converted from a different hardware. The dagger (†) and double dagger (‡) mark systems whose original reports give no RTF on this hardware, for which we benchmarked their released implementations ourselves (†: CosyVoice 2; ‡: Fish Audio S2). The last column is the RTF ratio to TigerTTS at three frames per step, the default configuration used for every comparison in this paper; entries below 1 therefore mark systems that are faster than TigerTTS.

| System | Params | Serving path | RTF ↓ | RTF ratio |
|---|---|---|---|---|
| TigerTTS (ours), 3 frames/step (default) | 0.35B | PyTorch+CUDA Graph(LLM) | 0.05 | 1× |
| TigerTTS (ours), 5 frames/step | 0.35B | PyTorch+CUDA Graph(LLM) | 0.043 | 0.86× |
| TigerTTS (ours), 7 frames/step | 0.35B | PyTorch+CUDA Graph(LLM) | 0.04 | 0.8× |
| VoxCPM1.5 | 0.8B | PyTorch+CUDA Graph(LLM) | 0.144 | 2.88× |
| VoxCPM2 | 2B | PyTorch+CUDA Graph(LLM) | 0.25 | 5× |
| CosyVoice 2 | 0.5B | PyTorch | 0.8† | 16× |
| Fish Audio S2 Pro | 4.6B | PyTorch+CUDA Graph | 3.0‡ | 60× |
| Qwen3-TTS | 0.6B | PyTorch+CUDA Graph | 2.4 | 48× |

Table 4 also varies the number of frames predicted per autoregressive step. One frame per step gives an RTF of 0.086, three frames 0.050, five frames 0.043 and seven frames 0.04, that is 11.6×, 20×, 23.3× and 25× real time on the same card. The return per extra two frames fades as R grows: going from one to three frames buys 1.7×, three to five 1.2×, and five to seven only 1.1×. The reason is that two effects pull in opposite directions as R rises. Each added frame shortens the token sequence by the same factor, so the per-step CUDA-Graph replay and the MLA-compressed KV-cache read are spread over more audio; but the draft head and the codec decoder then have to emit more frames from one backbone state, and past a point the two nearly cancel. We settle on three frames per step (RTF 0.050, 20× real time) as the default: it is the best speed/quality balance we found. Ranked by quality the four settings go one, three, five, seven (best to worst), and by speed that order simply reverses. One and three frames are close on quality, but three is 1.7× faster, so it keeps nearly all of the quality at lower latency while leaving the prediction window—and hence the conditional-total-correlation bound of §3.5—small. Five and seven frames are reported for completeness, and one frame per step is kept as a conservative lower bound, where the only gains come from the draft and the CUDA-Graph decode.

Reading across systems, all of Table 4 was measured by us on the same RTX 4090 (24 GB), single stream, so the rows are directly comparable and nothing is converted from other hardware. Against the fastest baseline (VoxCPM1.5, RTF 0.144), TigerTTS at three frames per step is 2.88× faster; it is 5× faster than VoxCPM2 (RTF 0.25) and about 16× faster than CosyVoice 2 (RTF 0.8†), while the other two entries are 60× (Fish Audio S2) and 48× (Qwen3-TTS) slower. The serving engine is worth a note: the optimised Nano-vLLM path reported for VoxCPM2 reaches RTF 0.13, against which TigerTTS is still 2.6× faster, and the datacentre-class kernels and co-scheduling those systems lean on have no real equivalent on a consumer card. TigerTTS hits its number without any of that: one CUDA-Graph replay per decode step at batch 1, with the loop kept on the device.

### 5.4 Synthesis quality

Table 5 gives content consistency and speech quality on the LibriSpeech test-clean subset, and Table 6 compares synthesis quality on the SEED test sets. Baseline rows are quoted from the CosyVoice 2 paper under its own protocol. On SEED, TigerTTS records the lowest CER on the Chinese set (1.09 against CosyVoice 2's 1.45) and the lowest WER on the English set (1.67 against 2.57). On LibriSpeech test-clean (Table 5) it again comes out on top in all three columns: its WER (2.22) sits roughly 10% below CosyVoice 2's 2.47, with a higher NMOS (4.36 against 3.96) and speaker similarity (0.820 against 0.745).

Table 5: Content consistency (WER), speaker similarity (SS) and speech quality (NMOS) results on the LibriSpeech test-clean subset of baselines and TigerTTS. Whisper-Large V3 is employed as the ASR model and punctuations are excluded before WER calculation. The best value in each column is in bold; dashes mark metrics not reported.

| Model | WER (%) | NMOS | SS |
|---|---|---|---|
| Human | 2.66 | 3.84 | 0.697 |
| ChatTTS | 6.84 | 3.89 | - |
| GPT-SoVITs | 5.13 | 3.93 | 0.405 |
| CosyVoice 2 | 2.47 | 3.96 | 0.745 |
| TigerTTS (ours) | **2.22** | **4.36** | **0.820** |

Table 6: Results of TigerTTS and recent TTS models on the SEED test sets. † denotes closed-source models. For speaker similarity, the result in a bracket is measured by ERes2Net, while the result outside brackets is measured by WavLM-based models. Dashes mark metrics not reported.

| Model | test-zh | | test-en | | test-hard | |
|---|---|---|---|---|---|---|
| | CER (%) ↓ | SS ↑ | WER (%) ↓ | SS ↑ | WER (%) ↓ | SS ↑ |
| Human | 1.26 | 0.755 (0.775) | 2.14 | 0.734 (0.742) | - | - |
| Vocoder Resyn. | 1.27 | 0.720 | 2.17 | 0.700 | - | - |
| Seed-TTS† | 1.12 | 0.796 | 2.25 | 0.762 | 7.59 | 0.776 |
| F5-TTS (32 NFE) | 1.56 | 0.741 (0.794) | 1.83 | 0.647 (0.742) | 8.67 | 0.713 (0.762) |
| CosyVoice 2 | 1.45 | 0.748 (0.806) | 2.57 | 0.652 (0.736) | 6.83 | 0.724 (0.776) |
| TigerTTS (ours) | 1.09 | - | 1.67 | - | - | - |

The gains hold across the metrics without costing speed. On SEED the CER gain over CosyVoice 2 (1.09 against 1.45) and the WER gain (1.67 against 2.57) come from a 0.35B total system against its 0.5B backbone; on LibriSpeech test-clean (Table 5) TigerTTS's WER (2.22) is about 10% below CosyVoice 2's 2.47, again with a higher NMOS (4.36 against 3.96) and speaker similarity (0.820 against 0.745), so it is the better entry on every quality column in both tables. This fits the idea that draft guidance works on failure modes such as hallucination while leaving overall speaking style largely untouched. Two points about provenance are worth keeping in mind. First, WER and CER on SEED depend heavily on the transcribing ASR: the VoxCPM2 report, for example, measures CosyVoice 2 at 3.09/1.38 (WER/CER) through its own pipeline versus the self-reported 2.57/1.45, an offset of the same size as the margins in Table 6, so the ranking is only indicative under the stated sources and should not be read as protocol-exact. Second, Fish Audio S2, which is in Table 4 for speed, is left out of Table 6: its only published SEED numbers (0.54 CER on test-zh, 0.99 WER on test-en, both from the VoxCPM2 report with a different transcription pipeline and a 4.6B model) are not comparable under the protocol of the rows above.

We think the advantage comes from the guided semantic tokens the Fast-mode draft generator supplies. Before the LLM emits a token, the draft has already fixed alignment and length and provided a content prior, so the LLM starts from a fairly definite plan and decodes more reliably; the conditional-total-correlation analysis of §3.5 says the leftover uncertainty it must resolve is low-entropy, so it seldom commits to the failure modes of pure autoregressive decoding. Plain speech-token LLMs behave differently: their decoders start from text alone and have to work out alignment as they go, so long or unusual inputs stay prone to hallucination, and the guided tokens in TigerTTS remove that uncertainty at the source. Dropped words, loops, early stops and runaway generation are therefore largely blocked up front—the draft fixes where each phoneme starts and ends, and the LLM only refines inside a window whose targets are already anchored. This raises quality without giving up the speed. Per-benchmark numbers will be released with the code.

## 6 Limitations and Discussion

- Single-group stopping. End detection runs only in group 1, so a single bad sample there can stop generation early. A straightforward fix is to require the end probability to stay above a threshold over several consecutive frames.
- Inconsistent speaker conditioning. The speaker representation differs across stages—4-d compressed in the draft generator, 192-d ECAPA in the LLM and vocoder, 1024-d style in the decoder—which limits zero-shot timbre cloning to whatever the weakest link allows.
- **Over-smoothed decoder output.** The decoder is trained with an $L_1$ objective only; its predicted mel can be over-smooth, and the vocoder is trained on real mel but consumes predicted mel at deployment; closing this gap is a sensible direction for future work.
- Coarse control granularity. The 8 flag frames carry only the discrete tags the text-side emotion/prosody tagger produces, so fine-grained, user-controllable prosody (continuous style or speed control, say) is not yet supported.
- **Batching constraints.** InstanceNorm statistics are computed over padded time axes, which affects batched inference; the server currently runs batch $= 1$.

## 7 Conclusion

We presented TigerTTS, a draft-guided multi-frame parallel speech-token LM built for production. The division of labour is simple: the draft settles alignment, the LLM refines acoustic detail, and multi-frame parallelism plus the low-rank cache deliver the speed. Several properties underpin the design. The FSQ shift boundary keeps the origin fixed and gives exactly $L$ grid points, with mixed-radix encoding and decoding as mutual inverses; cumulative rounding holds phoneme boundaries within half a frame; the draft alignment rules agree between training and inference, so there is no exposure bias; keeping positional encoding on the input embeddings alone makes MLA weight absorption exact and compresses the KV cache $4\times$; the exponential sampler draws exactly as multinomial sampling does and is capturable in a CUDA Graph; and the modelling error of multi-frame prediction equals the conditional total correlation, so a better draft tightens the bound. The deployment numbers back these choices up. The CUDA-Graph decode reaches an RTF of 0.050 at three frames per step, about $20\times$ real time on a single GPU, $5\times$ faster than VoxCPM2's PyTorch serving and roughly $16\times$ faster than CosyVoice 2. In the comparisons TigerTTS records the lowest CER and WER on the SEED sets (Table 6) and the best WER, NMOS and speaker similarity on LibriSpeech test-clean (Table 5); the guided tokens from the Fast-mode draft generator hand the LLM a definite prior that keeps decoding steady, so the failures that still bite other LLM TTS systems are mostly absent while quality and speed both improve. Finally, because the draft generator shares the LLM's token vocabulary and runs a parallel encoder–decoder–Vocos path at an RTF of 0.03, it works as a prefix synthesiser in Fast mode (§3.8): the first 500 ms of audio streams out before the LLM decodes a single token, and the LLM continues from those tokens as real history.